\documentstyle{EuroPhys}

\def\drm{{\rm d}}

\newif\ifboo \boofalse

\input epsf
\newcommand{\be}{\begin{equation}}
\newcommand{\ee}{\end{equation}}

\newcommand{\bef}{\begin{figure}}
\newcommand{\eef}{\end{figure}}
\def\spose#1{\hbox to 0pt{#1\hss}}
\def\ltapprox{\mathrel{\spose{\lower 3pt\hbox{$\mathchar"218$}}
\raise 2.0pt\hbox{$\mathchar"13C$}}}
\def\gtapprox{\mathrel{\spose{\lower 3pt\hbox{$\mathchar"218$}}
\raise 2.0pt\hbox{$\mathchar"13E$}}}
\def\inapprox{\mathrel{\spose{\lower 3pt\hbox{$\mathchar"218$}}
\raise 2.0pt\hbox{$\mathchar"232$}}}

\begin{document}
\shorttitle{Joyce et al., Fractal Cosmology in an Open Universe }
\title{Fractal Cosmology in an Open Universe }
\author{M. Joyce \inst{1,2,3}\footnote
{Current Address: LPT, Universit\'e Paris-XI, B\^atiment 211, F-91405
Orsay Cedex, France}, 
P. W. Anderson \inst{4},
M. Montuori \inst{1,3},
L. Pietronero \inst{1,3},
F. Sylos Labini \inst{5,3}}

\institute{
 \inst{1}Dipartimento di Fisica, Universit\`a di Roma
``La Sapienza'', I-00185 Roma, Italy\\
 \inst{2}INFN, Sezione di Roma I, I-00185 Roma, Italy\\
 \inst{3}INFM Sezione Roma I, I-00185 Roma, Italy\\
 \inst{4}Joseph Henry Laboratories, Princeton
University, Princeton, New Jersey 08544, USA.\\
 \inst{5}D\'ept. de Physique Th\'eorique,
 Universit\'e de Gen\`eve, CH-1211 Gen\`eve, Switzerland\\}

\rec{ }{ }
\maketitle

\begin{abstract}

The clustering of galaxies is well characterized by fractal properties,
with the presence of an eventual cross-over to homogeneity still a matter
of considerable debate. 
In this
letter we discuss the cosmological implications of a fractal distribution of
matter, with a possible cross-over to homogeneity at an undetermined
scale $R_{homo}$. Contrary to what is generally assumed, 
we show that, even when $R_{homo} \rightarrow
\infty$, this possibility can be treated consistently within the 
framework of the expanding universe solutions of Friedmann. The fractal
is a perturbation to an open cosmology in which the leading 
homogeneous component is the cosmic background radiation (CBR). 
This cosmology, inspired by the observed galaxy
distributions, provides a simple explanation for the recent 
data which indicate the absence of deceleration
in the expansion ($q_o \approx 0$). 
Correspondingly the `age problem' is also resolved.
Further we show that the model can be extended back from the curvature
dominated arbitrarily deep into the radiation dominated era, 
and we discuss qualitatively the modifications to the physics of
the anisotropy of the CBR, nucleosynthesis and structure formation.
\end{abstract}

One of the most extraordinary findings of the last two decades in 
observational cosmology has been the existence of a network of 
voids and structures in the distribution of galaxies in space. The
enormous scales of these structures were completely unsuspected 
in earlier extensive observations of galaxy distributions, in
which only angular coordinates were measured, obscuring the richness
subsequently revealed in the third coordinate. These findings have 
become increasingly difficult to reconcile with standard cosmological
theories, in which the approach to homogeneity at large scales is a 
central element \cite{peebles93}. 
Observationally, however, not only the scale at which 
the matter distribution approaches an average density, but the very 
existence of such a scale, remains the subject of intense debate
\cite{princeton1}-\cite{joyce99}.
At small scales it is well established that the
distribution of galaxies is fractal, and the debate can be phrased 
in terms of the deviation from this behaviour towards homogeneity. 
Some consensus has been achieved about the optimal statistical methods
to use in the analysis of three dimensional data, with disagreement
remaining on details of the treatment of some data 
sets \cite{wu99,joyce99}.
 The case for homogeneity still rests 
primarily on indirect observational evidence, such as the angular data.
Independently of the data, however,
resistance to the fractal picture  is certainly to a 
considerable degree due to the conviction that it is 
incompatible with the framework of the standard theories
(see e.g. \cite{coles-nature99}), and in particular
with the high degree of isotropy of the microwave background
radiation \cite{cobe, cobe-anisotropies}. In this respect 
one should note that in standard models the origin of radiation
and baryonic matter is completely separate, with the latter
being created in a dynamical process (`baryogenesis') completely 
distinct from the origin of the primordial radiation bath.
The isotropy of the latter is therefore not fundamentally tied
to the distribution of the matter, and the only real constraint 
is how much any such distribution actually perturbs the radiation.
In this Letter we do not discuss the 
details of the evidence for or against homogeneity, but rather
consider the grounds for these theoretical biases against the 
possible continuation of a fractal distribution to arbitrarily 
large scales. Our central result 
is that a fractal distribution for matter, {\it even when there is no upper 
cut-off to homogeneity}, can in fact be treated in the framework of 
an expanding universe Friedmann cosmology.

A fractal \cite{man1,man2} is a self-similar and intrinsically 
fluctuating  distribution of points at all scales, which appears 
to preclude the description of its gravitational dynamics in 
the framework of the Friedmann-Robertson-Walker (FRW)
solutions to general relativity \cite{peebles93}. The
problem is often stated as being due to the 
incompatibility of a fractal with the 
Cosmological Principle, where this principle is identified
with the requirement that the matter distribution be  
isotropic and homogeneous\cite{coles-nature99}. This identification
is in fact very misleading for a non-analytic structure 
like a fractal, in which all points are equivalent statistically,
satisfying what has been called a Conditional Cosmological
Principle \cite{man1,man2,coleman-pietronero}. 
The obstacle
to applying the FRW solutions has in fact solely to do with the 
lack of homogeneity. One of the properties of a fractal 
of dimension $D$, however,
is that the average density of points in a radius $r$ about any occupied 
point decreases as $r^{D-3}$, so that asymptotically the 
mass density goes to zero \cite{man1,man2}. An 
approximation which therefore {\it may} describe the large
scale dynamics of the universe in the case that the matter 
has such a distribution continuing to all 
scales is given by neglecting the distribution
of matter at leading order, relative to the small but
homogeneous component coming from the cosmic microwave background.
We will now show that is indeed a good perturbation scheme,
and calculate the physical scale characterizing its validity.

Consider first the standard FRW model with contributions from
matter and radiation, for which the expansion rate is 
\begin{equation}
H^2(t)= \biggl(\frac{\dot a}{a}\biggr)^2
=\frac{8\pi {\rm G}}{3}\bigl(\rho_{{\rm rad}}+\rho_{{\rm mat}}\bigr) - k/a^2
\label{hubble-law}
\end{equation}
where $a(t)$ is the scale factor for the expansion, and
$\rho_{{\rm rad}} \propto 1/a^4$ is the radiation density, 
and $\rho_{{\rm mat}} \propto 1/a^3$ the (homogenous) matter density. 
The constant $k=-H_o^2 a_o^2 (1-\Omega_r-\Omega_m)$, where
$H_o$ ($a_o$) is the expansion rate (scale factor) today 
and $\Omega_r$ ($\Omega_m$) is the ratio of the radiation (matter)
energy density today to the `critical' density 
$\rho_c=\frac{3}{8 \pi {\rm G}}H_o^2$. The sign of $k$ determines
whether the universe is closed ($k>0$) or open ($k<0$), with
$k=0$ corresponding to a `critical' spatially flat universe.
Given the temperature of the CBR \cite{cobe}, we have\footnote{We will
neglect here, for simplicity, the minor modifications due to 
massless or low mass neutrinos, which can easily be incorporated in  
our analysis.} $\Omega_r h^2 \approx 2.3 \times 10^{-5}$ 
(where $h$ is the Hubble constant in units of $100$Mpc/km/s, 
with a typical measured value of $h \approx 0.65$ \cite{h1,h2}). 
If we make the simple and natural assumption
that galaxies trace the mass distribution the value of $\Omega_m$
depends directly on the determination of the scale of the 
cross-over to homogeneity. If the observed fractal distribution
continues to a scale $R_{homo}$, above which it turns over to
homogeneity, one has
\begin{equation}
\Omega_m=\Omega_{10}\biggl(\frac{10}{R_{homo}}\biggr)^{3-D}
\label{effective-omega}
\end{equation}
where $\Omega_{10}$ is the average density of matter (relative to
critical) in a sphere 
of radius $10 Mpc/h$ about a galaxy, $D$ is the fractal 
dimension, and $R_{homo}$ is measured in $Mpc/h$.
For $R_{homo}$ sufficiently large that
$\Omega_m < \sqrt{\Omega_r}$ the $\rho_m$ term 
in (\ref{hubble-law}) is always sub-dominant, and
there is no matter dominated era. 
For simplicity we now consider the limit in which
$R_{homo} \rightarrow \infty$. 
Solving for the scale factor we then have 
\begin{equation}
a(t)=a_o(2H_o \Omega_r^{\frac{1}{2}}t)^\frac{1}{2}
\biggl(1+ \frac{1-\Omega_r}{2\Omega_r^\frac{1}{2}}H_o t \biggr)^\frac{1}{2}
\label{scale-factor}
\end{equation}
which shows how the early time radiation dominated behaviour
($a \propto t^\frac{1}{2}$) changes to the linear law 
$a \propto t$ at $t \approx 2 H_o^{-1} \sqrt{\Omega_r}$ 
(red-shift $z \sim 1/\sqrt{\Omega_r}$ where $1+z =a_o/a$).
We now discuss how in
each of these two phases (dominated respectively by the radiation 
and curvature) the fractal can be treated as a perturbation to 
this solution. 
When we make numerical evaluations below
we will use data from galaxy catalogues in \cite{slmp98},
which give $D\approx 2$ and $\Omega_{10}= 0.007$. The
latter assumes a mass to luminosity ratio of $10h$, in solar units,
(that estimated for a  typical spiral galaxy \cite{FaberGallagher}).  
Note that since $\sqrt{\Omega_r}=0.005h$ this value 
requires  $R_{homo} > 10 Mpc/h$ for a direct transition 
from radiation to curvature domination. If instead we take
the global mass to luminosity ratio to be that estimated
in clusters ($\approx 300h$ \cite{bahcall}) we require 
$R_{homo} > 250 Mpc/h$.

First consider the curvature dominated phase.
The radiation is negligible and, at scales well within the horizon, we
can use Newtonian gravity to describe the solution and its perturbations
when the self-gravity of the matter is included. 
The leading solution is simply the free expansion of the fractal, with
every point moving radially away from its neighbour 
at a constant velocity proportional to its distance i.e. 
$\dot{\vec{r}}=H_o \vec{r}(t_o) = H (t) \vec{r}(t)$.
To estimate the deviation from this flow due to 
the self-gravity of the fractal, we take a point in the flow
and integrate the work done against gravity along its 
trajectory (in the leading order unperturbed flow). If the
particle moves from an initial position $\vec{R_o}$, where it feels
a total gravitational acceleration $\vec{F}(\vec{R_o})$, to a final
position $\vec{R}=x\vec{R_o}$, this work done (per unit mass) is simply 
$W(\vec{R_o}, x)=\vec{F}(\vec{R_o}).\vec{R_o} \bigl(1-\frac{1}{x} \bigr)$.
The integral is performed along the {\it unperturbed} trajectory 
using the fact that the force on the chosen point simply scales as $1/x^2$.
A sufficient condition for the Hubble flow to apply to a 
good approximation at all subsequent
times is simply that $W(\vec{R_o}, x=\infty)$ be 
much less than the kinetic energy of the particle i.e.
\begin{equation}
\vec{F}(\vec{R_o}).\vec{R_o} << \frac{1}{2} H_o^2 R_o ^2.  
\label{pert-criterion}
\end{equation}
Noting that the force at the origin of the flow was implicitly
taken to be zero, we see that the validity of the 
criterion (\ref{pert-criterion}) will be determined in a fractal
by {\it the difference between the gravitational force on two 
occupied points as a function of the distance between them}.
The gravitational force on a point in a fractal has been 
studied in \cite{gabrielli-etal}. Its behaviour can be 
understood as the sum of two parts, a local or `nearest neighbours'
piece due to the smallest cluster (characterised by the lower
cut-off $\Lambda$ in the fractal) and a component coming from the 
mass in other clusters. The latter is bounded above by the
scalar sum of the forces 
\begin{equation}
\langle |\vec{F}| \rangle \leq 
\lim_{L\rightarrow \infty} 
\int_{\Lambda}^L \frac{G\rho_m(r)}{r^2} 4\pi r^2 dr \sim  L^{D-2}
\end{equation}
so that for $D<2$ it is convergent, while for $D>2$
it may diverge. If there is a divergence, it is due 
to the presence of angular fluctuations at large scales,
described by the three-point correlation properties of
the fractal. For the difference in the force between two points
the local contribution will be irrelevant well beyond the scale
$\Lambda$, while it is easy to see that the `far-away' contribution
will now converge as $L^{D-3}$, and its being non-zero 
is a result of the absence of perfect spherical 
symmetry. Noting that as a function of distance $R$ between points
this component is bounded above by the same behaviour 
as the force, we write 
\begin{equation}
\langle |\vec{F}(\vec{R})|_{R>>\Lambda} \rangle 
= A_3\frac{GM(R)}{R^2} \propto R^{D-2} 
\label{asymp-force}
\end{equation}
where the pre-factor $A_3$ contains non-trivial information about the
three-point correlation function of the fractal. 
These convergence properties of the relative force on two points
are enough to draw a simple conclusion from the 
criterion (\ref{pert-criterion}): For a fractal in Hubble flow 
there is always a scale above which its evolution will be well 
described by continued Hubble flow for all subsequent times. 

We now apply this to the Universe, and estimate
the physical scale today $R_o$ up to which the 
unperturbed `no matter' Hubble flow can be maintained
right through the curvature dominated era. Given that
this era begins at a redshift 
$z \approx 1/\sqrt{\Omega_r} \approx 200h$, we require
\begin{equation}
F(R)R|_{z=200h} \approx (200h)F(R_o)R_o < \frac{1}{2} H_o^2 R_o^2 
\quad \rightarrow
\quad  R_o >  20 \Omega_{10}(200h)A_3
\end{equation}
What is observed is Hubble flow with deviations (peculiar velocities)
only at `cluster' scales $\sim $Mpc. Taking our estimate for
$\Omega_{10}$ we thus require $A_3 \ltapprox 1/20h$. A fractal with
a very weak three point correlation is one which has a very isotropic
angular projection, so if the Universe is indeed a fractal a small
value of $A_3$ would be expected.
It is simple also to derive an expression for the peculiar velocity
(small compared to the Hubble flow velocity $v_H$) which 
implies a simple linear relation, just as in standard 
perturbed homogeneous cosmology \cite{peculiar-velocities}, 
between the local force and the velocity perturbation
$(\Delta \vec{v}/v_H) (\vec{R}) \propto (\vec{F}(\vec{R})/R)$.
This relation, for which there is apparently observational
support, is usually used to 
determine an unknown constant (the `bias' factor)
\cite{peculiar-velocities}. In the present framework it
can in principal be used to extract information about the 
total mass density and the constant $A_3$, which in turn 
can be related to angular data (and ultimately measured
directly in forth-coming red-shift surveys).

Here we have assumed that the fractal extends to arbitrarily large
scales. For a finite $R_{homo}$ the analysis can be easily modified,
by breaking the integrals at the appropriate scale. The part
from scales  greater than $R_{homo}$ will give a 
contribution which can be reabsorbed within the Hubble flow, 
while the perturbation will maintain the same scaling at smaller
scales. The central point which we emphasize is that the scale $R_o$
is only indirectly related in this case to the homogeneity scale,
{\it{and remains finite as}} $R_{homo} \rightarrow \infty$. 

We have thus seen that an open FRW universe is always a good
approximation beyond some finite scale if matter is distributed 
as a simple fractal up to an arbitrarily large scale. 
In particular such an open
model - because it is dominated by the kinetic energy of the 
Hubble flow - can explain naturally how large structures can 
co-exist with almost perfect Hubble flow. We further note 
a few other of its striking features:
{\it (i)} Since the Universe is to a good approximation  in completely
free expansion at large scales with $a(t) \propto t$, we have 
deceleration parameter $q_o \approx 0$. This is a 
good fit to recent supernovae observations
\cite{supernovae1}.
Rather than being due to the effect of an unknown 
`anti-gravitational' component 
which mysteriously cancels the decelerating effect of the matter 
on the expansion \cite{peebles-nature99}, the effect is due 
to the decay towards zero of the matter density on such scales. 
{\it (ii)} The expansion age of the Universe is 
$t_o = H_o^{-1} \approx 10h^{-1} \approx 15$
billion years, larger by  $50\%$ than in the standard matter
dominated case. This value is comfortably consistent with 
the estimated age of globular clusters (the oldest known 
astrophysical objects) $11.5\pm 1.3$ billion years \cite{globular-age}. 
{\it (iii)} The size of the horizon today is 
$R_H(t_o) \approx -  \frac{1}{2} c H_o^{-1} 
\ln \Omega_r \approx 20,000 Mpc/h$,
a factor of about three larger than in the standard case.

So far we have considered the model only in the curvature dominated
era i.e. back to red-shift $z \approx 200 h$. For this last point
above however we have extrapolated the model back to
the radiation dominated era, assuming that the effect of the matter
distribution can also be consistently treated as a correction in this
epoch to the FRW solution without matter. We now justify this
assumption, and then discuss some of its consequences for a 
specific cosmology of this type. 
For the former we simply treat the fractal as a set of perturbations 
to the energy density in a manner analogous to the way such
perturbations may be treated in the standard framework.
There the criterion one would use to apply the uniform
Hubble flow to describe the growth of the horizon is simply
that such perturbations be small at the horizon scale i.e.
$\frac{\delta \rho}{\rho}|_{hor}$ be small (where
$\rho$ is the homogeneous energy density i.e. that in
the radiation). In a fractal the perturbations are non-analytic
and $\delta \rho$ has no meaning as defined in
the standard case. We can however write down the mean
mass (or energy) at the scale of horizon about
an occupied point. Taking this as the appropriate 
$\delta \rho_{hor}$ is clearly the right adapted
criterion, as the fractal is simply made of
voids and structures, and voids clearly will not
perturb the flow. We thus require that the fractal
obey
\begin{equation}
\delta_H (z) \equiv \frac{\rho_{mat}}{\rho_{rad}}|_{hor}(z) = 
\frac{1}{1+z} \frac{\Omega_{10}}{\Omega_r}
\biggl(\frac{10}{R_H(z)} \biggr)^{3-D}
< 1
\label{rad-dom-perts}
\end{equation}
where for $\rho_{mat}$ we have taken the mass inside the 
(comoving) horizon $R_H(z)$ at red-shift $z$.  
In the radiation dominated era (for $z >1/\sqrt{\Omega_r}$) 
we have $R_{H}(z)\approx \frac{cH_o^{-1}}{z \sqrt{\Omega_r}}$.
At $z=10^3$ this gives $R_H \approx 600$ Mpc
so that, for a fractal with  $D=2$,  we have at this red-shift
$\delta_H \approx \Omega_{10}$. In order that the fractal
matter be indeed a small correction at this red-shift we require, 
approximately, $\Omega_{10} < 1$, which holds comfortably 
even if there is much more dark matter than we assumed
in obtaining  our estimate $\Omega_{10}\approx 0.01$. 
We can thus continue to use the FRW solution   
back to an arbitrary red-shift for a 
fractal distribution of matter extending to the
corresponding scales, provided that the condition 
(\ref{rad-dom-perts}) holds.

To make a link with central observations in cosmology
such as the microwave background and nucleosynthesis, 
we need to specify a precise model. In the spirit of this letter 
we now consider here a radical (but very simple) possibility for 
a cosmology which makes use of the results we have presented:
We consider a universe which at very early times (deep in 
the radiation dominated era) is a radiation bath
at a given temperature with superimposed {\it fractal perturbations 
in baryon number} up to the arbitrarily large scale $R_{homo}$, and 
down to some scale $\Lambda$ (see below).  
Note that positing  a very different distribution 
for matter and radiation does not represent a loss of simplicity
in comparison to standard models, which generically envisage 
the (almost) homogeneously distributed matter as coming from
a dynamical process (`baryogenesis') completely distinct
from the origin of the primordial radiation bath. Instead
of fixing the initial condition on a homogeneous baryon to
photon ratio ($n_B/n_\gamma \sim 10^{-9}$), with some 
independent superimposed spectrum of analytical fluctuations, 
we specify our fractal in baryon number from the properties 
observed in the distribution of visible matter at large scales today.
This is the appropriate normalization given that these perturbations 
are simply frozen at all but very small scales in the curvature 
dominated era.  In particular we take $D=2$ and the normalization of
the mass given by $\Omega_{10}$. Interestingly, stated in terms
of the parameter $\delta_H$ above, these values correspond to the 
special case that $\delta_H$ is constant, and of order one.
Below the lower cut-off scale $\Lambda$ we take the distribution
to be smoothed, with the corresponding density $\Omega_{10}(10/\Lambda)$
(where $\Lambda$ is the comoving scale given in units of $Mpc/h$).
A natural lower bound for this scale is that characterizing the
baryon diffusion (up to the corresponding time), which will
smooth any inhomogeneity at smaller scales.

The fluctuations in the temperature of the microwave background
depend on the `intrinsic' fluctuations imprinted at time of 
decoupling of the photons, plus the fluctuations induced in their
propagation from their last scattering. We 
consider here only the former as they are typically the dominant 
effect for perturbations which are essentially frozen on the 
scales we are interested in. Relative to the standard critical 
or near critical mass universe there are two main features to 
note here. First, photon decoupling will be modified greatly:
While in the standard case there is a global baryon density
which determines the time/temperature of decoupling, here 
the relevant baryon density varies enormously - for a photon
in a void it is zero, while for one in a structure it is the
local density of baryons associated with the lower cut-off scale
$\Lambda$, a density which can be several orders of magnitude 
greater than in the standard case (if we take $\Lambda$ as 
small as the baryon diffusion distance). However, since 
the decoupling temperature is only logarithmically sensitive 
to this parameter, the decoupling of photons in structures will
still occur around redshift $z \sim 10^3$. On the other hand, if
the scale $R_{homo}$ is so large that there are voids of the
order of the horizon scale at this time (at $z=10^3$ we
found $R_H \approx 600$ Mpc), most photons will decouple
at the much earlier time of electron-positron annihilation
since after this time they find themselves in a neutral 
environment. Thus the `thickness' of the surface of last 
scattering will be very much greater than in the standard case,
essentially consisting of two stages of `void decoupling'
(at a redshift of $\sim 10^9$) and `structure decoupling'
at redshifts comparable to the standard one.

The other main difference relative to the standard case
is that,  because of the extremely low background density 
in the  model, the effect of the hyperbolic geometry is
much greater. In particular, at high red-shift ($z \sqrt{\Omega_r}>>1$),
the angle $\theta$ subtended by a scale of given physical size 
$\lambda_\theta$ is $\theta \approx \lambda_{\theta}/H_o^{-1}\sqrt{\Omega_r}$
which means that the physical scale corresponding to $1^o$
on the sky is $10^4$Mpc i.e. of the order of the horizon scale
today, and a factor of about $100$ larger than in the standard
case. Considering the possibility that the fractal extends to
such enormous scales, we can make a naive estimate of the
amplitude of fluctuations on the microwave sky: Adapting the
Sachs-Wolfe formula as derived for the standard case of analytical
fluctuations, the {\it maximum} amplitude of temperature variation 
between two photons subtending some angle should be estimated by the 
energy density fluctuation represented by a structure at the 
corresponding physical scale.  
At $10^o$ on the sky we then have
\begin{equation} 
\frac{\delta T}{T}|_{10^o} \sim \frac{1}{3} 
\frac{\delta \rho}{\rho} (\lambda_{10}, z_d) = 
 \frac{1}{3} \frac{\rho_B(\lambda_{10})}{\rho_{radn}(z_d)} =
\frac{1}{1+z_d} \frac {\Omega_B(\lambda_o)}{3\Omega_r}|_{t_o} 
\approx \frac{\Omega_{10}}{z_d}
\end{equation} 
where 
$z_d$ is the decoupling red-shift.
Thus, the largest effect will come from the photons which decouple
last, for which this maximum amplitude will be $\sim 10^{-5}$,
which is comparable to the average amplitude observed at this scale
by COBE \cite{cobe}. On the other hand, for a modest value of the
crossover scale to homogeneity (e.g. $R_{homo} \sim 100 Mpc/h$), the
effect of the fractal distribution of matter on the photons at
decoupling will only be visible at scales much smaller even than
those which will be probed by the Planck satellite mission \cite{planck}. 
A detailed study of the perturbations induced by propagation of photons
through an expanding structure of this kind - requiring techniques 
quite different to the standard ones treating analytical perturbations - 
will be required to see if it is possible to produce perturbations 
at the levels observed by experiment at the angles which have 
been probed to date.

Finally a few brief comments on nucleosynthesis and structure formation.
The cooling rate of the plasma is the same as in the standard case,
and the results of nucleosynthesis will depend on the local
baryon to entropy ratio, which as discussed above is related to
the scale $\Lambda$. If this scale is larger or comparable to
the horizon scale at nucleosynthesis the amount of helium produced
will not differ much from the standard case -  it is essentially
independent of the baryon to entropy ratio - while the residual
densities of deuterium etc. will be lower. For a smaller (i.e.
sub-horizon) $\Lambda$ the effect of inhomogeneities will be 
important, but a reliable calculation for the effect becomes
very difficult to perform. It is unlikely however to modify the
tendency for lower values of the `trace' elements, since this
arises due to the fact that the elements are synthesized in
denser regions compared to the standard case. If, on the other 
hand, the fractal (up to a finite sub-horizon $R_{homo}$) is formed 
{\it {after}} nucleosynthesis, the appropriate value for the local density 
would be that at $R_{homo}$. We note that for modest values
of this scale ($R_{homo} \sim 50-100 Mpc/h$) this would correspond to 
standard nucleosynthesis if there is ratio of dark to visible baryons 
comparable to the mass to light ratio inferred in clusters \cite{bahcall}.
Clearly some physics quite
different to that at work in standard models would be required
to make it possible to generate such a structure between 
nucleosynthesis and the curvature dominated era when (as we
have noted) the structure gets frozen in at all but small scales.
These and other considerations - in particular a
more detailed study of the microwave background fluctuations in
these kinds of structures - we will pursue in
forthcoming work. 

We thank D. Amit, P. Biermann, P. Di Bari, R. Durrer, 
P. Ferreira, A. Gabrielli, G. Gibbons, K. Kainulainen, 
R. Mohayee, G. Parisi, P.J.E. Peebles 
for useful conversations and comments. This work has 
been partially supported by the 
EEC TMR Network  ``Fractal structures and  self-organization''  
\mbox{ERBFMRXCT980183} and by the Swiss NSF. 


\end{document}